\input{aipcheck}

\documentclass[
    ,final            
  ]
  {aipproc}

\layoutstyle{8x11single}

\usepackage{amsmath,amssymb,graphicx}
\def\ignore#1{{}}

\def\mydate{8 June 2012}
\def\onehalf{{\frac{1}{2}}}
\def\onethird{{\frac{1}{3}}}
\def\twothird{{\frac{2}{3}}}
\def\eff{{\rm eff}}

\def\go{{\rightarrow}}

\def\beeq{\begin{equation}}
\def\eneq{\end{equation}}
\def\beqn{\begin{eqnarray}}
\def\eeqn{\end{eqnarray}}

\def\mybig{\displaystyle \strut }
\def\mbig{\displaystyle }
\def\myfrac#1#2{{\mybig #1\over \mybig #2}}

\def\la{\raise.16ex\hbox{$\langle$}\lower.16ex\hbox{}  }
\def\ra{\raise.16ex\hbox{$\rangle$}\lower.16ex\hbox{} }
\def\Tr{{\rm Tr}\,}

\begin{document}

\title{Gauge-Higgs Unification Approach}

\classification{11.10.Kk, 11.15.Ex, 12.60.-i}
\keywords      {symmetry breaking, gauge-Higgs unification, 
Hosotani mechanism, lattice simulations}

\author{Yutaka Hosotani}{
  address={Department of Physics, Osaka University\\
  Toyonaka, Osaka 560-0043, Japan}
}

\begin{abstract}
When the extra dimensional space is not simply-connected, 
dynamics of the AB phase in the extra dimension can induce dynamical 
gauge symmetry breaking by the Hosotani mechanism.  This opens up
a new way of achieving unification of gauge forces.  It leads to
the gauge-Higgs unification.  The Hosotani mechanism
can be established nonperturbatively by lattice simulations,
in which measurements of the Polyakov line give a clue.
\hfill (OU-HET 751/2012, \mydate)
\end{abstract}

\maketitle


\vskip -10pt
\centerline{\small \it To appear in the Proceedings of GUT 2012, Kyoto, 15-17 March 2012.}

\subsection{I. Introduction}

Unification of gauge forces is achieved by starting with larger symmetry 
at high energies  than the directly observed symmetry at low energies.  
The symmetry is spontaneously  broken, being reduced to the observed one.  
The symmetry breaking mechanism constitutes the backbone of the unification.  

There are various ways to achieve spontaneous gauge-symmetry breaking.
The most popular one is the Higgs mechanism.  A scalar field develops
a nonvanishing vacuum expectation value at the tree level which is not invariant
under gauge transformations.  In the Coleman-Weinberg mechanism 
radiative corrections to the effective potential of the respected scalar fields
induce symmetry breaking in a theory with
no dimensionful parameter at the tree level.  In technicolor theory
scalar condensates of fermion-antifermion pairs break the symmetry.

In higher dimensional gauge theory there appears another way of
breaking gauge symmetry.  Extra-dimensional components of gauge
potentials, as a consequence of dynamics, can induces gauge symmetry
breaking.  It is called the Hosotani mechanism.

\subsection{II. Hosotani mechanism}

When the extra-dimensional space is not simply connected, there appears
an Aharonov-Bohm (AB) phase along a non-contractible loop in the extra
dimension.  This AB phase is a part of physical degrees of freedom
of gauge fields.   Its value is dynamically determined.  With a nontrivial value 
it leads to gauge symmetry breaking.\cite{YH1, Davies1, Hatanaka1998}

\subsection{(a) QCD at $\mathbf {T \not= 0}$ v.s.  on $\mathbf {S^1}$}

The idea of the Hosotani mechanism came by examining QCD 
at finite temperature ($T \not= 0$).
Finite temperature field theory (for equilibrium) is equivalent to  field
theory with an imaginary time $\tau$ in  an interval $(0, \beta = 1/k_B T)$
with boundary conditions that 
all bosonic (fermionic) fields are periodic (anti-periodic).
The imaginary time has  topology of $S^1$. 
Gluons in $SU(N_c)$ gauge theory acquire effective masses at finite temperature 
$m^2 = \onethird g^2 T^2 (N_c + \onehalf N_F)$  where $N_F$ is the number 
of fermions in the fundamental representation.
Quark-gluon plasma at $T \not= 0$ gives screening of gluon propagation.

Now consider QCD on $R^1 ({\rm time}) \times R^2 \times S^1 ({\rm space})$
where one spatial dimension is a circle $S^1$ with a circumference $\beta$.   
After Wick rotation of the time
axis the theory is the same as QCD at $T \not=0$ except that boundary 
conditions become less restrictive.  One can impose a boundary condition
$\psi(x, y+ \beta) = e^{i\delta} \psi(x, y)$ for fermions.  
Nothing is wrong with imposing a periodic boundary condition $e^{i\delta} = 1$.
It is an easy exercise to show that the effective gluon masses are changed:
\beeq
m^2 = \onethird g^2 T^2 (N_c + \onehalf N_F)
\quad \go \quad
m^2 = \onethird g^2 T^2 (N_c -  N_F) ~.
\label{mass1}
\eneq
What happens if $N_F > N_c$?  Does   $m^2 < 0$ imply the instability of 
the vacuum?  It turns out  that $\la A_y \ra \not= 0$ in the true vacuum and
$\la A_y \ra \not= 0$ can lead to gauge symmetry breaking in non-Abelian gauge theory.

\subsection{(b) Dynamics of AB phases}

Consider a gauge theory on a product of   $d$-dimensional Minkowski spacetime
$M^d$ and a circle $S^1$ with a coordinate $y$ and radius $R$.  
Finite temperature QCD in 4D corresponds to the $d=3$ case.
The relevant quantities for the vacuum structure 
are  Aharonov-Bohm (AB) phases  along $S^1$:
\beeq
W = P \exp \Big\{ i g \int_C dy \, A_y \Big\} ~.
\label{ABphase1}
\eneq
In $SU(N)$ gauge theory eigenvalues of $W$ are given by 
$\{ e^{i\theta_1}, \cdots,  e^{i\theta_N} \}$,  $\sum_{j=1}^N \theta_j = 0 ~(mod ~ 2\pi)$.  
Note that constant $A_y$ is nontrivial.  
Eigenvalues of $W$ are gauge invariant so that they  cannot be gauged away.
Even if $\theta_j$'s give vanishing field strengths, they represent physical degrees of
freedom, affecting physics at the quantum level.  They are AB phases. 
If $e^{i\theta_j } \not= e^{i\theta_k }$, it leads to gauge symmetry breaking.

The values of $\theta_j$'s are not at our disposal.  They are dynamically determined,
once the matter content in the theory is specified.  The true vacuum corresponds to
the global minimum of the effective potential $V_\eff (\theta_j)$.
At the tree level $V_\eff (\theta_j)^{\rm tree} = 0$, as field strengths vanish.
At the quantum level it becomes nontrivial.  Particles in $M^d$ 
consist of Kaluza-Klein towers, the spectra of  which typically take the form of 
$m_n(\theta_H) = R^{-1} \big( n + \theta_H /2\pi \big)$ ($n$: an integer).  
Here $\theta_H$ represents AB phases $\theta_j$ collectively.  The spectrum
depend on $\theta_H$.  The effective potential at 1-loop is
\beeq
V_\eff (\theta_H)^{\rm 1 \, loop}
= \sum \pm \onehalf \int \frac{d^d p}{(2\pi)^d} \sum_n \ln \Big\{ 
p^2 + m_n(\theta_H)^2 \Big\} ~.
\label{effV1}
\eneq
It is remarkable that the $\theta_H$-dependent part of $V_\eff (\theta_H)$ is  finite, 
being free of divergence for any $d$.\cite{YH1, Hosotani2005, HMTY2007}
As a consequence the global minimum of $V_\eff (\theta_H)$ is unambiguously 
determined.  

\subsection{III. Gauge-Higgs unification}

This opens up a new way of having dynamical gauge symmetry breaking.  
Dynamics of AB phases in extra dimensions can induce gauge symmetry breaking.
Four-dimensional fluctuations of these phases correspond to 4D Higgs fields.
The gauge-Higgs unification is achieved.  Higgs fields are identified with a part
of extra-dimensional components of gauge potentials.  

\subsection{(a) $\mathbf {SU(3)}$ on $\mathbf M^d \times S^1$}

Consider  $SU(N)$ gauge theory on $M^d \times S^1$.  Let us suppose that 
all fields are periodic on $S^1$.  In terms of $\theta_j$ ($j=1, \cdots, N$) 
spectra of massless particles are given by
\beeq
m_n = 
\begin{cases}
\mbig \frac{1}{R} \Big( n + \frac{\theta_j - \theta_k}{2\pi}  \Big)
&{\rm for~adjoint~rep.} , \cr 
\noalign{\kern 5pt}
\mbig \frac{1}{R} \Big( n + \frac{\theta_j }{2\pi}  \Big)
&{\rm for~fundamental ~rep.}
\end{cases}
\label{spectrum1}
\eneq
Once the matter content is specified, $V_\eff (\theta_H)$ in (\ref{effV1}) is
evaluated.  
For $d=4$ the  effective potential is given by
\beqn
&&\hskip -1cm
V_\eff(\theta) = C \Bigg\{
-3 \sum_{j,k=1}^N h_5 \bigg( \myfrac{\theta_j - \theta_k}{2\pi} \bigg) 
+ 4 N_{\rm fund}^F 
\sum_{j=1}^N h_5 \bigg( \myfrac{\theta_j - \beta_{\rm fund}}{2\pi} \bigg) 
+ 4 N_{\rm ad}^F  \sum_{j,k=1}^N h_5 
\bigg( \myfrac{\theta_j - \theta_k- \beta_{\rm ad}}{2\pi} \bigg) 
 ~ \Bigg\} ~~, \cr
\noalign{\kern 10pt}
&&\hskip 2.cm
h_d(x) = \sum_{n=1}^\infty \myfrac{\cos 2\pi n x}{n^d} ~, 
\quad
C = \myfrac{3}{4\pi^2} \myfrac{1}{(2\pi R)^4} ~.
\label{exampleV1}
\eeqn
Here  $N_{\rm fund}^F$ and $N_{\rm ad}^F$ are the numbers of fermion
multiplets in the fundamental and adjoint representations, respectively.
$\beta_{\rm fund}$ and $\beta_{\rm ad}$ are the boundary condition parameters
appearing in $\psi(x, y + 2 \pi R)  =  e^{i\beta} \psi(x,  y)$.
In general, each multiplet of fermions can have distinct $\beta$.

It is tempting to apply this to  GUT, as GUT symmetry is
normally broken to the SM symmetry by a Higgs field in the adjoint representation.
For instance, if 
$e^{i\theta_1} = e^{i\theta_2} = e^{i\theta_3} \not= e^{i\theta_4} = e^{i\theta_5}$
is realized in $SU(5)$ theory, then one obtains 
$SU(5) \go SU(3) \times SU(2) \times U(1)$  gauge symmetry breaking.   
It turns out that  $SU(5)$ symmetry remains unbroken if fermion multiplets 
come in only {\bf 5} and {\bf 10}  with $\beta=0$.

\begin{figure}[h,t,b]
\centering  \leavevmode
\includegraphics[height=5.0cm]{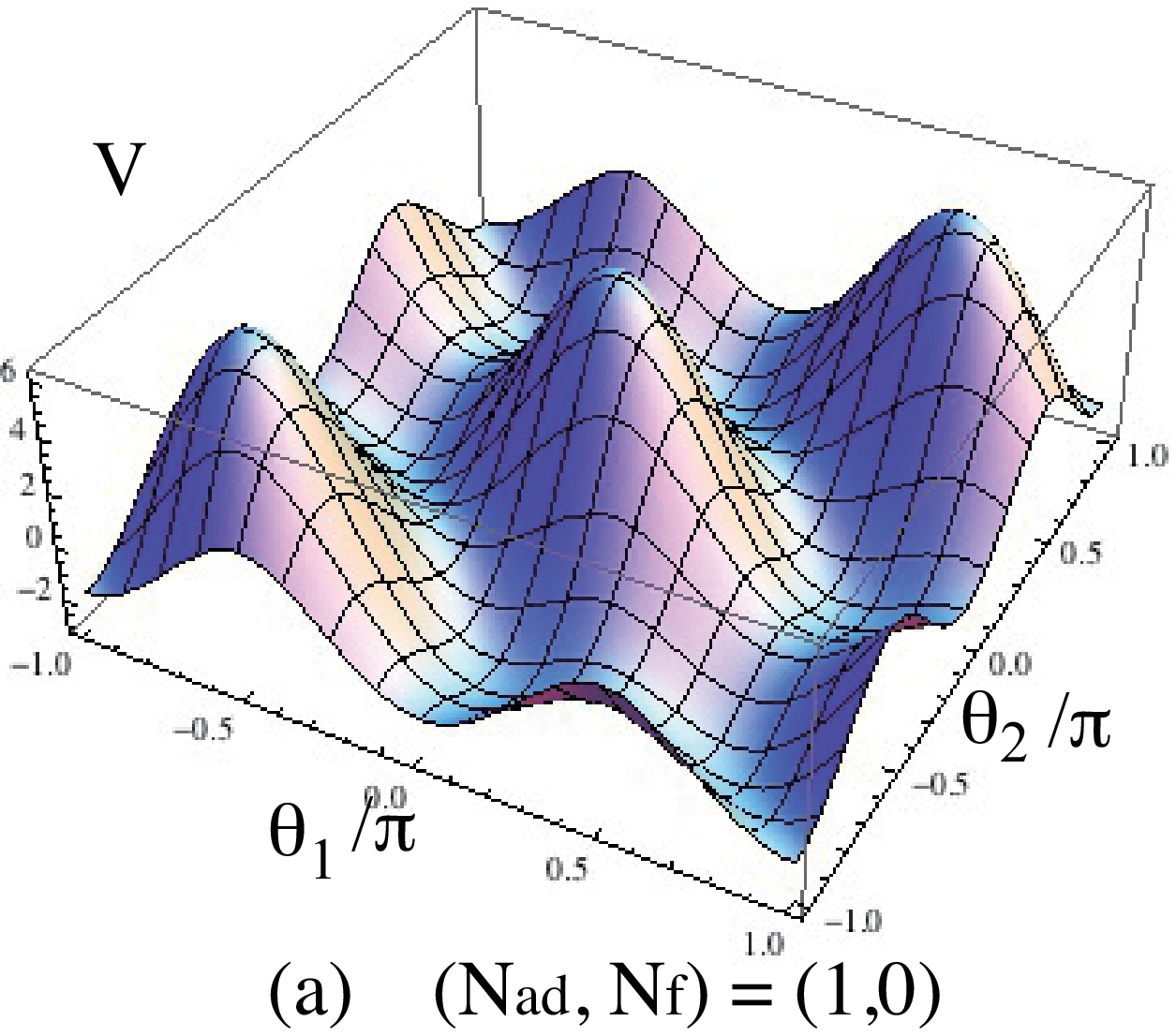}
\hskip 1.cm 
\includegraphics[height=5.0cm]{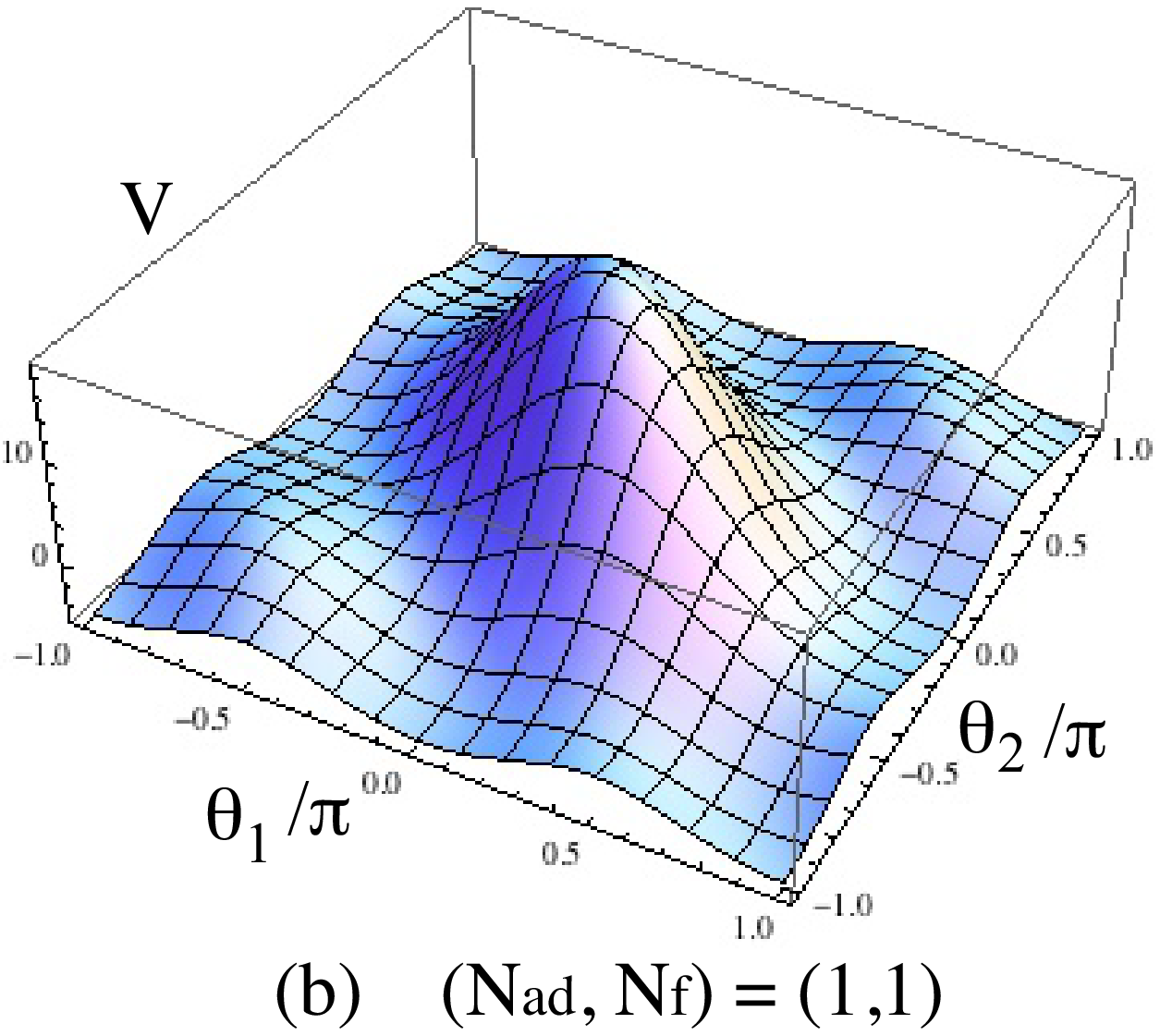} 
\caption{The effective potential  $V_\eff (\theta_1, \theta_2)$ in $SU(3)$ gauge 
theory on $M^4 \times S^1$ with massless fermions.  
(a) With $N^F_{\rm ad}=1, N^F_{\rm fund}=0$.
(b) With $N^F_{\rm ad}=N^F_{\rm fund}=1$.   
For both cases $\beta_{\rm ad} = \beta_{\rm fund} = 0$.
The  symmetry is broken to (a) $U(1) \times U(1)$, and (b) $SU(2) \times U(1)$.}
\label{effV-su3-fig1}
\end{figure}

Nontrivial examples of symmetry breaking are found when there are fermions 
in the adjoint representation.\cite{Davies1, Hosotani2005}
The effective potential  $V_\eff (\theta_1, \theta_2)$ in  $SU(3)$ gauge theory 
are displayed in Figure~\ref{effV-su3-fig1}.
In pure gauge theory $SU(3)$ symmetry is unbroken.  When one adjoint fermion is added
$(N^F_{\rm ad}, N^F_{\rm fund})=(1,0)$, $V_\eff$ is minimized at 
$(\theta_1, \theta_2, \theta_3) = (0, \frac{2}{3}\pi, -\frac{2}{3}\pi)$ and its
permutations.  The symmetry is broken to $U(1) \times U(1)$.  
When $(N^F_{\rm ad}, N^F_{\rm fund})=(1,1)$, $V_\eff$ is minimized at 
$(\theta_1, \theta_2, \theta_3) = (0, \pi, \pi)$ and its permutations.
The symmetry is broken to $SU(2) \times U(1)$.  

We note that the boundary conditions are $SU(3)$ symmetric in these examples.
Dynamics of the AB phases induce symmetry breaking.

\vskip 10pt

\subsection{(b) Electroweak unification and GUT}

We have seen that non-Abelian gauge symmetry can be dynamically broken by the 
Hosotani mechanism.  It is interesting to apply this mechanism to electroweak
unification and GUT.  To have realistic models one has to incorporate chiral fermions,
which becomes highly nontrivial in higher dimensional gauge theory. 
One powerful way to have chiral fermions is to consider models in which extra-dimensional
space is an orbifold.  The simplest example of orbifolds is $S^1/Z_2$.  Two points
$y$ and $-y$ on $S^1$ are identified.  There appear two fixed points at $y=0$ and 
$\pi R$ on $S^1$, which are customarily called as two branes.  

Many years ago Pomarol and Quiros formulated the standard model on 
$M^4 \times (S^1/Z_2)$.\cite{Pomarol1} 
Since then many models have been 
proposed.\cite{Burdman2003}-\cite{HosotaniMoriond}
With intensive  experiments going on at LHC, which report possible 
candidates for the Higgs boson, it becomes
necessary to make definitive predictions to be tested.

The most promising model of gauge-Higgs unification for electroweak interactions is
the $SO(5) \times U(1)_X$ model in the Randall-Sundrum warped 
space.\cite{ACP}-\cite{HOOS}
$SO(5) \times U(1)_X$ breaks down to $SO(4) \times U(1)_X$ by the orbifold
boundary conditions, to $SU(2)_L' \times U(1)_Y'$ by brane dynamics, and 
to $U(1)_{\rm EM}$ by the Hosotani mechanism.  It has been shown that the 
dynamical EW symmetry breaking takes place thanks to the presence of the top 
quark.  The most striking result is that the Higgs boson, with a mass predicted 
around 130 GeV, becomes absolutely stable.\cite{HKT, HTU1}
The effective potential  $V_\eff (\theta_H)$ is minimized 
at $\theta_H = \onehalf \pi$.  There emerges new parity ($H$-parity) under which 
the Higgs boson is odd, while all other SM particles are even.

Historically the Hosotani mechanism was first applied to GUT models.\cite{YH1}
On orbifolds the doublet-triplet splitting problem in $SU(5)$ GUT 
can be naturally solved.\cite{Kawamura2000}
Having chiral fermions and GUT symmetry breaking by the Hosotani mechanism 
simultaneously, however,  is nontrivial.\cite{Hall1, Lim1}  
It is also known that boundary conditions
at the fixed points of orbifolds fall into equivalence classes. \cite{HHHK,  HHK}
Apparently different
boundary conditions lead to the same physics as a consequence of dynamics
of AB phases if those boundary conditions belong to the same equivalence class.
New proposals for GUT have been made in Ref. \cite{Yamashita2011}.

\subsection{IV. Nonperturbative Hosotani mechanism}

So far the Hosotani mechanism for dynamical gauge symmetry breaking has
been established only in perturbation theory.  It is important to establish it
nonperturbatively.  We would like to describe  how to do it on lattice.  

There have already been lattice studies which, as explained below, support
the Hosotani mechanism.\cite{Irges}-\cite{Debbio}  Myers and Ogilvie studied $SU(3)$ and 
$SU(4)$ gauge theories  at finite temperature with periodic boundary conditions
for fermions.   Depending on fermion content, they claimed
that there appear new phases.\cite{Myers}  
Cossu and D'Elia investigated $SU(3)$ gauge theory 
on $16^3 \times 4$ lattice with massive fermions in the adjoint representation, 
examining Polyakov lines.  They found phase transitions separating
``confined'', ``deconfined'', ``split'' and ``re-confined'' phases.\cite{Cossu}
We show that all these results can be understood well with the notion of
the Hosotani mechanism.

\subsection{(a) Polyakov line and $\mathbf V_{\bf eff}$}

Let us consider $SU(3)$ gauge theory on $M^3 \times S^1$ with massive fermions.
The Wilson line $W$ in (\ref{ABphase1}) along $S^1$ has three eigenvalues,
$(e^{i\theta_1},  e^{i\theta_2},  e^{i\theta_3})$ where $\sum \theta_j =0$.
The Polyakov line is 
\beeq
P = \frac{1}{3} \, \Tr W = \frac{1}{3} \sum_{j=1}^3 e^{i\theta_j} 
= \frac{1}{3} \Big\{ e^{i\theta_1} + e^{i\theta_2} + e^{-i (\theta_1 +\theta_2) } \Big\} ~.
\label{Pline1}
\eneq
In lattice simulations $\la P \ra$ is measured, which includes all quantum fluctuations.
Corresponding to $M^3 \times S^1$, the simulations are done on 
$N_1^3 \times N_2$ lattice ($16^3 \times 4$ in ref.~\cite{Cossu}).

The parameters in the continuum theory are $g$ (gauge coupling), $R$ 
(the radius of $S^1$), and $m$ (fermion mass).  Those in the lattice theory are
$\beta$ (lattice gauge coupling), $a$ (lattice spacing) and $a m$.
The fermion mass $m$ plays an important role in the lattice simulation.
In the continuum theory $V_\eff$ at  the one loop level takes the form
$V_\eff^{\rm 1 \, loop} = g^2 R^{-3} f(\theta_1, \theta_2, \kappa)$ where
$\kappa = \pi R m$.

$V_\eff (\theta_1, \theta_2)$ is nontrivial.  In the strong coupling regime (in the 
confinement phase), or for sufficiently large $R$, quantum fluctuations are
large.  All values of $(\theta_1, \theta_2)$ are almost equally taken so that
$\la P \ra = 0$.  In the lattice simulations $\la P \ra$ should be centered around
the origin in the complex plane.

\begin{figure}[h,t,b]
\centering  \leavevmode
\includegraphics[height=4.8cm]{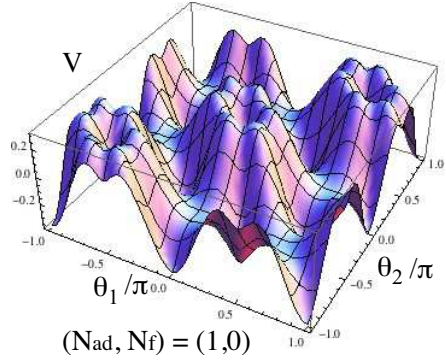}
\caption{The effective potential  $V_\eff (\theta_1, \theta_2)$ in $SU(3)$ gauge 
theory with one adjoint fermion with  $\kappa =\pi R m=0.55$.
}
\label{effV-su3-fig2}
\end{figure}

In the weak coupling regime dominant gauge configurations are localized
around one of the minima of $V_\eff (\theta_1, \theta_2)$.
Here the fermion mass as well as the gauge coupling becomes
important.  The method to evaluate  $V_\eff$ with massive fermions has 
been developed in Ref.~\cite{HH2011}.  $V_\eff^{\rm 1\, loop}$ with one
adjoint fermion with a mass $\kappa = \pi R m=0.55$ is depicted in 
Fig.~\ref{effV-su3-fig2}.  Notice that there develops more structure in
$V_\eff$ than in the massless case in Fig.~\ref{effV-su3-fig1}.

\vskip 10pt

The wave function of the AB phases  $\theta_1, \theta_2$ has finite spreading.  
It implies that the magnitude $| \la P \ra |$ get smaller
than the value evaluated at the minimum of $V_\eff$.  In $SU(3)$ theory
the phase of $\la P \ra$ are nontrivial and one can see a transition from one
minimum to another.   This is exactly what has been observed in the lattice
simulations.

\subsection{(b) Classification of phases}

Examination of $V_\eff^{\rm 1 \, loop}$ with a given fermion mass
shows that minima of $V_\eff$ are always located at some specific 
points.  This behavior seems to persist to all order in perturbation theory,
and seems to be supported by lattice simulations.

Phases classified by the Polyakov line are listed in Table~\ref{table1}.
In each category the values of $(\theta_1, \theta_2, \theta_3)$ can
take permutations of the given value.  In the phase $C$, for instance,
there are six degenerate global minima of $V_\eff$.
$(A_1, A_2, A_3)$ and $(B_1, B_2, B_3)$ form $Z_3$ multiplets.
If there is no fermion in the fundamental representation, then
the three phases in each $Z_3$ multiplet are degenerate.

\vskip 13pt
\begin{table}[h,t,b]
\begin{tabular}{c|c|c|c|c}
\hline
Phase& $(\theta_1, \theta_2, \theta_3)$ & $P$ & Symmetry 
& Names used in ref.~\cite{Cossu}\\
\hline
$X$ & large fluctuations & 0 & $SU(3)$ & confined \\
\hline
$A_1$, $A_{2,3}$ 
   & $(0,0,0)$, $(\pm \twothird \pi, \pm \twothird \pi,\pm \twothird \pi)$
   &$1, \, e^{\pm 2 i \pi  /3}$ &$SU(3)$ & deconfined\\
\hline
$B_1, B_{2,3}$
  &$(\pi, \pi, 0)$, $(\pm \onethird \pi, \pm \onethird \pi, \mp \twothird \pi)$
  &$- \onethird, \, \onethird e^{\pm i\pi/3}$
  &$SU(2) \times U(1)$ &split\\
\hline
$C$ &$(0, \twothird \pi, - \twothird \pi)$
  &0  & $U(1) \times U(1)$ &re-confined \\
\hline
\end{tabular}
\caption{Classification of various phases in $SU(3)$ gauge theory.
Location of the minima of $V_\eff$, Polyakov line $P$, and residual symmetry
are listed.}
\label{table1}
\end{table}
\vskip 7pt

\begin{figure}[b,h]
\centering  \leavevmode
\includegraphics[height=4.cm]{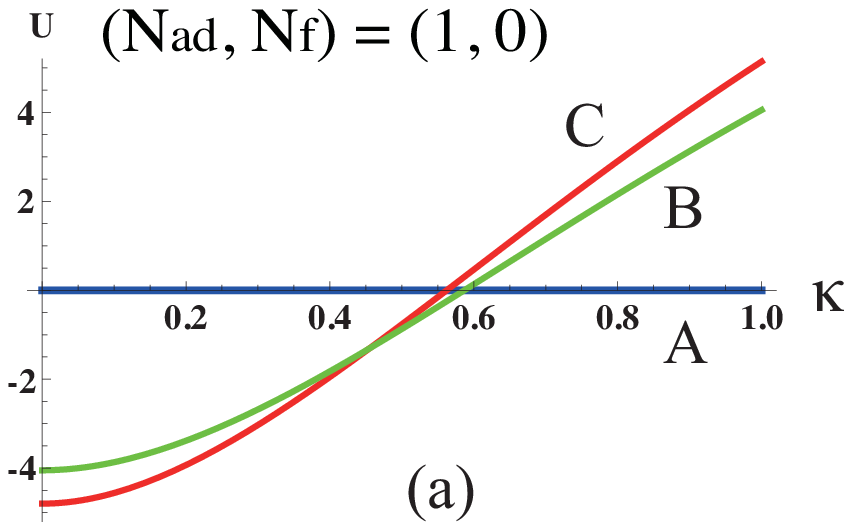}
\hskip .7cm
\includegraphics[height=4.cm]{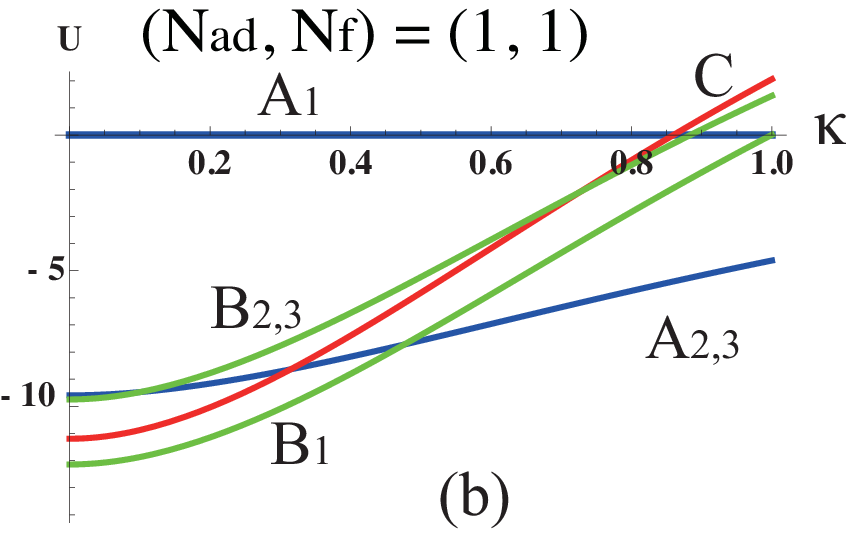}
\caption{The value of the effective potential  $V_\eff^{\rm 1 \, loop} $ 
in various phases  in $SU(3)$ gauge  theory as a function of   $\kappa =\pi R m$. 
(a) $(N_{\rm ad}, N_{\rm fund}) = (1,0)$.
(b) $(N_{\rm ad}, N_{\rm fund}) = (1,1)$.
}
\label{effV-su3-fig3}
\end{figure}

In lattice simulations the values of the coupling $\beta$ and the fermion mass
times lattice spacing  $ma$ are varied.  
$V_\eff$ at the 1 loop in the continuum theory 
is evaluated  with varying $\kappa= \pi R m$.  
We have plotted  the values of $V_\eff $ in various phases in 
Fig.~\ref{effV-su3-fig3}.   
One can infer the pattern of the phase transitions:
\beqn
&&\hskip -1.cm
{\rm Case~ (a)} \quad (N_{\rm ad}, N_{\rm fund}) = (1,0) : \qquad
X ~~ \Leftrightarrow ~~ A ~~ \Leftrightarrow  ~~ B 
~~  \Leftrightarrow  ~~C \cr
\noalign{\kern 5pt}
&&\hskip -1.cm
{\rm Case~ (b)} \quad (N_{\rm ad}, N_{\rm fund}) = (1,1) : \qquad
X ~~ \Leftrightarrow  ~~ A_{2,3}  ~~ \Leftrightarrow ~~ B_1  ~.
\label{phase2}
\eeqn
Cossu and D'Elia have observed the same transition pattern in their
lattice simulation.\cite{Cossu}  Apparently large $\kappa$ corresponds to
large $R$, which in turn corresponds to small $|V_\eff |$.
Fluctuations due to gauge interactions become more important, 
and therefore it effectively corresponds to large gauge coupling.  
More investigation is necessary to pin down the phase structure. 

\vskip 10pt 

\subsection{V.  Summary}

The above result indicates that dynamical gauge symmetry breaking
by the Hosotani mechanism is taking place in $SU(3)$ gauge theory on 
$M^3 \times S^1$ when there are fermions in the adjoint representation.
The lattice studies are mostly done in four dimensions to avoid the convergence
issue in higher dimensions.  
 If the Hosotani mechanism works in higher dimensions nonperturbatively,
it gives a new paradigm for unifying gauge forces.


\subsection{Acknowledgments}

I would like to thank Etsuko Itou and Jim Hetrick for many enlightening
comments on the lattice results, and Hisaki Hatanaka for clarifying phase 
transitions in the presence of massive fermions.
This work was supported in part 
by  scientific grants from the Ministry of Education and Science, 
Grants No.\ 20244028, No.\ 23104009 and  No.\ 21244036.

\bibliographystyle{aipproc}   


\begin{thebibliography}{99}

\bibitem{YH1}
Y.~Hosotani,
\Journal{\PLB}{126}{309}{1983}; 
\Journal{\AP}{190}{233}{1989}.

\bibitem{Davies1}
A.~T.~Davies and A.~McLachlan,
\Journal{\PLB}{200}{305}{1988};
\Journal{\NPB}{317}{237}{1989}.

\bibitem{Hatanaka1998}
H.\ Hatanaka, T.\ Inami and C.S.\ Lim,
\Journal{\MPLA}{13}{2601}{1998}.


\bibitem{Hosotani2005}
Y.~Hosotani,
Proceeding for SCGT2004, arXiv: hep-ph/0504272. 
 
\ignore{
\bibitem{Hosotani2006}
Y.~Hosotani,
arXiv: hep-ph/0602146.
}

\bibitem{HMTY2007}
Y.~Hosotani, N.~Maru, K.~Takenaga and T.~Yamashita,
\Journal{\PTP}{118}{1053}{2007}.


\bibitem{Pomarol1}
A.\ Pomarol and M.\ Quiros, \Journal{\PLB}{438}{255}{1998}.

\bibitem{Burdman2003}
G.\ Burdman and Y.\ Nomura, \Journal{\NPB}{656}{3}{2003}.

\bibitem{Csaki2003}
C.\ Csaki, C.\ Grojean and H.\ Murayama, \Journal{\PRD}{67}{085012}{2003}.


\bibitem{ACP}
K.~Agashe, R.~Contino and A.~Pomarol,
\Journal{\NPB}{719}{165}{2005}.

\bibitem{MSW}
A.~D.~Medina, N.~R.~Shah and C.~E.~M.~Wagner, 
\Journal{\PRD}{76}{095010}{2007}.

\bibitem{HS2} 
Y.~Hosotani and Y.~Sakamura,
\Journal{\PTP}{118}{935}{2007}.

\bibitem{HOOS} 
Y.~Hosotani, K.~Oda, T.~Ohnuma and Y.~Sakamura,
\Journal{\PRD}{78}{096002}{2008}.  
{\it Erratum}, \Journal{\ibid}{79}{079902}{2009}.

\bibitem{Giudice}
G.F.~Giudice, C.~Grojean, A.~Pomarol and R.~Rattazzi,
\Journal{\JHEP}{0706}{045}{2007}.


\bibitem{HKT}
Y.~Hosotani, P.~Ko and M.~Tanaka,
\Journal{\PLB}{680}{179}{2009}.


\bibitem{HNU}
Y.\ Hosotani, S.\ Noda and N.\ Uekusa,
\Journal{\PTP}{123}{757}{2010}.

\bibitem{HTU1}
Y.\ Hosotani, M.\ Tanaka and N.\ Uekusa,
\Journal{\PRD}{82}{115024}{2010}; 
\Journal{\PRD}{84}{075014}{2011}. 


\bibitem{Agashe2010}
K.~Agashe, A.~Azatov, T.~Han,  Y.~Li,  Z.-G.~Si, and L.~Zh,
\Journal{\PRD}{81}{096002}{2010}.


\bibitem{Contino2011}
R.~Contino, D.~Marzocca, D.~Pappadopulo and R.~Rattazzi,
\Journal{\JHEP}{1110}{081}{2011}.

\bibitem{HosotaniMoriond}
Y.~Hosotani, 
arXiv:1205.3837 [hep-ph].

\bibitem{Kawamura2000}
Y.~Kawamura, \Journal{\PTP}{103}{613}{2000}; 
\Journal{\PTP}{105}{999}{2001}.


\bibitem{Hall1}
L.\ Hall and Y.\ Nomura, 
\Journal{\PRD}{64}{055003}{2001}.

\bibitem{Lim1}
M.\ Kubo, C.S.\ Lim and H.\ Yamashita,
\Journal{\MPLA}{17}{2249}{2002}.

\bibitem{HHHK}
N.\ Haba, M.\ Harada, Y.\ Hosotani and Y.\ Kawamura, 
\Journal{\NPB}{657}{169}{2003};   
{\it Erratum}, {\it ibid.}  B{\bf 669} (2003) {381}.

\bibitem{HHK}
N.\ Haba,  Y.\ Hosotani and Y.\ Kawamura, 
\Journal{\PTP}{111}{265}{2004}.



\bibitem{Yamashita2011}
K.~Kojima, K.~Takenaga and T.~Yamashita,
\Journal{\PRD}{84}{051701}{2011};
T.~Yamashita,
\Journal{\PRD}{84}{115016}{2011}.

\bibitem{Irges}
N. Irges, K. Knechtli,
\Journal{\NPB}{775}{283}{2007}.


\bibitem{Myers}
J.C.~Myers and M.C.~Ogilvie,
\Journal{\PRD}{77}{125030}{2008};
arXiv:0809.3964 [hep-lat];
\Journal{\JHEP}{0907}{095}{2009}.

\bibitem{Cossu}
G.~Cossu and M.~D'Elia,
\Journal{\JHEP}{0907}{048}{2009}.

\bibitem{Forcrand}
P.\ de Forcrand, A.\ Kurkela, M.\ Panero,
\Journal{\JHEP}{1006}{050}{2010}.

\bibitem{Debbio}
L.~Del Debbio, E.~Rinaldi and A.~Hart, 
arXiv:1203.2116 [hep-lat].

\bibitem{HH2011}
H.\ Hatanaka and Y.\ Hosotani,  arXiv:1111.3756[hep-ph].



\end{thebibliography}

\def\Journal#1#2#3#4{{#1} {\bf #2}, #3 (#4)}

\def\NPB{{\em Nucl. Phys.} B}
\def\PLB{{\em Phys. Lett.}  B}
\def\PRL{\em Phys. Rev. Lett.}
\def\PRD{{\em Phys. Rev.} D}
\def\AP{{\em Ann.\ Phys.\ (N.Y.)} }
\def\MPLA{{\em Mod.\ Phys.\ Lett.} A}
\def\PTP{{\em Prog.\ Theoret.\ Phys.}}
\def\ibid{{\em ibid.} }
\def\JHEP{{\em JHEP} }

\end{document}